# ON SETTING OF HEAT-AND-MASS TRANSFER PROBLEMS UNDER DIRECTED CRYSTALLIZATION

A.P.Guskov

Institute of Solid State Physics RAS, Chernogolovka, Moscow Distr., Russia guskov@issp.ac.ru

## Introduction.

So far the problem of interface behavior upon phase transition has not yet acquired a satisfactory mathematical formulation due to a variety of the physical phenomena involved. Analytical solutions exist only for elementary problems describing the free interface behavior in directed crystallization conditions, for instance, for those implying a clearly shaped isothermal interface (ellipsoid, paraboloid, hyperboloid) [1,2]. Numerical calculations of the interface behavior also present significant difficulties since the instability of moving interface does not enable calculations by means of known algorithms. The mathematical description of the moving interface problem includes transfer equations for each phase with corresponding initial and boundary conditions which should be specified in each phase as well as on the interface. The general solution of this problem does not seem possible now, so quantitative analysis of phase transition is made after significant simplifications of the problem commonly reduced to the so-called quasi-equilibrium problem setting which has been used in a number of papers [2, 3, 4]. Quasi-equilibrium problem setting was used for calculation of numerous technological processes [2,4,5,6] that are successfully used in crystal growth, metallurgy and materials science. It was used to study the reasons for the interface instability during phase transition. However, the solutions of the quasi-equilibrium problem are, as a rule, inherently qualitative. For instance, attempts have been made to use the results of interface stability analysis during crystallization for description of periodic eutectic structure [7]. In [8,9] a linear analysis of interface stability under the directed crystallization was made with regard to the kinetics of particle condensation to the growing crystal surface and the kinetics of the non-equilibrium solution layer in front of the interface. The problem setting allowed for kinetic overcooling which was shown to be responsible for the interface instability, the latter in its turn can lead to spatial distortions as well as temporary fluctuations of the temperature and concentration of the components on the interface during phase transition. The analysis of interface movement dynamics made in [10] revealed the existence of modes responsible for the periodic distribution of the solution component concentration on the planar interface. It was assumed that the mode corresponds to eutectic structure formation. Numerical calculations have been made of the interface component distribution period as a function of the interface velocity for a quadratic dependence between the interface velocity and kinetic overcooling [11]. This coincided with the experimental dependences of the eutectic structure period on the interface velocity. These dependences are known to be well approximated by rate power -1/2. This result was used to obtain analytical dependences of the period of morphological interface instability on the interface velocity for different growth mechanisms [12,13]. The dependences obtained coincided with the experimental dependences of the eutectic structure period on the interface velocity [14]. However, this failed to explain decay of the liquid solution in front of the interface without analyze of the stability of the overcooled solution. The analysis made in [15] showed that the overcooled solution in the vicinity of the eutectic point is not stable. In [16] it was also shown that unstable solution may decay with a period equal to that of morphological interface instability.

Crystallization of two-component liquid solution is described by the system of heat conduction and diffusion equations for solid and liquid phases as well as the corresponding boundary conditions. The boundary crystallization problem is described in detail in a number of studies. For instance in [2], the heat conduction and diffusion equations should be fulfilled in each phase volume

$$\frac{\partial \bar{T}'_r}{\partial \tau_r} = \chi'_r \Delta \bar{T}'_r \quad \frac{\partial \bar{T}_r}{\partial \tau_r} = \chi_r \Delta \bar{T}_r \quad (1)$$

$$\frac{\partial \bar{C}_r}{\partial \tau_r} = D_r \Delta \bar{C}_r \quad (2)$$

Here the values related to the solid phase are denoted by the prime, index r indicates the dimensional quantities, the overline indicates that the functions are written in the lab coordinates, $T$ is the temperature, $C$ concentration, $\chi$ thermal diffusivity, $t$ time, $D$ diffusion coefficient. The diffusion coefficient of the solid phase is assumed to be equal to zero. The problem does not take into account the liquid convection and crystal anisotropy. The heat transfer is written as a heat conduction equation. The interface thickness is assumed to be zero and on the interface have to be fulfilled the temperature continuity condition

$$\bar{T}'_r\big|_{solid} = \bar{T}_r\big|_{liquid}, \quad (3)$$

and the heat flow continuity condition

$$\chi'_r \frac{\partial \bar{T}'_r}{\partial n}\bigg|_{solid} - \chi_r \frac{\partial \bar{T}_r}{\partial n}\bigg|_{liquid} = \varepsilon v_n, \quad (4)$$

Here, $v_n$ is the interface velocity along the normal to the interface toward the liquid phase. The mass flow condition should also be fulfilled

$$D_r \frac{\partial \bar{C}_r}{\partial \mathbf{n}}\bigg|_{liquid} = v_{nr} \cdot (\bar{C}_r' - \bar{C}_r)\mathbf{n}\bigg|_{liquid} \quad (5)$$

Here **n** is the direction of the normal to the interface surface. These boundary conditions express the general conservation laws between the contacting phases. In analysis of interface stability in the one-dimensional case the interface temperature and infinitely remote point temperatures are specified [2]. In case a multi-component solution is crystallized, the system contains diffusion equation (2). In the solid phase diffusion is normally neglected. For problems on infinite interval, the initial concentration of the liquid solution and the distribution coefficient are given. The concrete form of the boundary conditions depends on the crystallization conditions and the degree of the simplifying conditions that are used to obtain a practically convenient solutions.

The quasi-equilibrium problem setting is the most popular approximation for directed crystallization problems. The quasi-equilibrium setting implies that the interface temperature is equal to the equilibrium temperature of the phase transition with due allowance made for the effect of the interface curvature, i.e. the Gibbs–Thomson effect. The distribution ratio is assumed to be equilibrium. The temperature distribution is found from homogeneous heat conduction equations (1) and specified are the values of the temperature phase gradients on the interface rather than the temperature in infinitely remote points. Such problem setting enabled to obtain a number of simple and practically useful solutions of directed crystallization problems [2-4]. However, comparison of the calculated and experimental data reveals that the solutions obtained do not take into account all the conditions of phase transition that can have qualitative effects on the processes involved in the problems in question. We have recently published several papers in which the setting of the directed crystallization problem includes nonequilibrium processes affecting heat-and-mass transfer during phase transitions [8-13]. The quantitative agreement between the calculated data and the experimental dependences of the eutectic structure period on the interface velocity indicate that the problem setting in the above papers takes into consideration the conditions which can qualitatively affect the heat-and-mass transfer processes during phase transitions. We present a detailed discussion of the boundary conditions of the directed crystallization problem, a formulation of the model considering temperature fields of external sources, the mechanism of attachment of particles to the growing solid surface, the influence of interphase component absorption on the phase distribution ratio of the components as well as the calculation of the period of the morphological interface instability which is made with due regard of all the aforementioned conditions.

## Environmental Heat Exchange.

We introduce environmental heat transfer into the heat conduction equation for two reasons. First, it enables us to state a problem with specified temperature values in infinitely remote points, second, it allows for eliminating physically unrealizable solutions of the stationary problem. Such solutions may occur due to the fact that in interface stability problems material heating is frequently simulated by specified temperature gradients. Such boundary conditions can be accepted provided the aim is to show the feasibility of unstable stationary modes of interface movement. However, in the experiments a temperature field is usually formed in the material by way of environmental heat exchange or induction heating [5]. In controlled crystallization the material is heated above its melting temperature followed by liquid phase cooling. In the case of environmental heat exchange crystallization occurs in the area of the controlled temperature gradient which is formed between the heater and the cooling unit. The temperature field is formed by different heat exchange units. The measurement of the interface temperature gradient is a complicated technological problem. The gradient values are generally found by solving the corresponding heat conduction problem. A homogeneous heat conduction equation (1) is obtained if environmental heat exchange is neglected. In one of the phases the solution of the homogeneous equation exhibits a divergence away from the interface, i.e. it yields a physically unrealizable function. Yet, the solution has relatively simple analytical forms and allows understanding the processes occurring at the interface. The situation changes in case material heating is considered and the equation includes the terms describing the so-called internal source field (ISF). Hence, the problem becomes inhomogeneous and more complicated since the interface coordinate in the lab coordinate system depends on its stationary velocity, the external temperature field and the thermophysical parameters of the crystallized material which determine kinetic overcooling. In the general case, in the coordinates that are stable with respect to the moving bar, the ISF position is time and coordinate dependent. In the lab coordinate system, in which the bar moves at a specified constant rate, the temperature field is independent of time. In accordance with [17], we will formally specify the ISF by the exponential function in each phase. It is a convenient approximation since the exponent is the solution of the inhomogeneous heat conduction equation when the bar velocity is equal to zero. On the interface, which is at the origin of coordinates, the ISF should satisfy the condition of temperature and heat flow continuity, i.e. the external field is given by the temperatures of the heater and the cooler and the temperature gradient on the interface. The temperature on the interface is equal to that of the phase transition. This means that the distribution of temperature in the stationary bar can be specified using external heat sources by way of heat transfer through the thin bar surface. Since transverse heat transfer is not considered, the temperature distribution is one-dimensional. When the finite stationary interface velocity is given, the position of the interface is shifted with respect to the specified heat source density function. In [17] this shift was determined by the known interface temperature which was calculated by the specified dependence of interface velocity on kinetic overcooling. This means that the stationary problem with the given external temperature field involves an additional condition, namely, the position of the interface with respect to the lab coordinate system. Let-in of the ISF allows to set the liquid and solid phase temperatures in infinitely remote points as boundary conditions. The temperature of the bar in infinitely remote points is equal to that of the heater and the cooler. The interface temperature is determined by the preset interface velocity. Hence, the position of the interface changes with respect to the lab coordinate system and the condition required for its determination will be specified when solving the stationary problem. Such problem setting was briefly considered in [17]. The distribution of external temperature in the vicinity of each phase is set as a sum of the constant term and the exponential function. The temperature distributions in the phases are joined on the interface by the following conditions.

1. Temperature continuity in point z=0.
2. Heat flow continuity in point z=0.
3,4. Phase temperatures $T'_{inf}$ and $T_{inf}$ are set in infinitely remote points in the vicinity of the liquid and solid phases.
5. Temperature gradient is set in point z=0 $\phi 0$.
6. Temperature is set in point z=0.

The function satisfying the first five conditions has the form

$$T_{ext}(z) = \begin{cases} T'_{inf} + \left(T_{inf} + x - T'_{inf}\right) \exp\left(\dfrac{\phi_0 z}{T_{inf} + x - T'_{inf}}\right) \\ T_{inf} + x \exp\left(\dfrac{\phi_0 z}{x}\right) \end{cases} \quad (6)$$

Here *x* is an unknown parameter to be found on solving the stationary problem.

**Consideration of solid phase growth kinetics.**

Let us consider the value of interface temperature. In the quasi-equilibrium problem setting the interface temperature is assumed to be equal to equilibrium temperature of phase transition. However, from statistical physics it is known that for phase transition particles have to overcome a potential barrier, so transition occurs at a temperature which is different from equilibrium phase transition temperature. This temperature difference is known as kinetic overcooling. The interface velocity is a function of kinetic overcooling. The function form depends on the growth mechanism. If the temperature of the interface is equal to that of the phase transition, the kinetic overcooling is zero. In this case, in accordance with the thermodynamics law, any growth mechanism yields a moveless interface. If the problem involves equal interface and phase transition temperatures, the velocity is independent of the phase transition non-equilibrium grain. Then the goal is to find the movement of the geometric surface where the temperature is equal to that of the phase transition, rather than the interface. If kinetic overcooling is neglected, then linearization of the problem involving a moving interface disregards the linear approximation of the interface velocity related to the growth kinetics. And it should be noted that the interface temperature in the stationary mode practically coincides with the equilibrium temperature of the phase transition. This follows from the limiting transition when the kinetic coefficient tends to zero as it is shown in [3,4].

Consider the expression for interface temperature and the terms of the linear expansion of the boundary-value problem which are lost if no account is taken of the dependence of the interface velocity on kinetic overcooling. Let us write the interface velocity as a function of kinetic overcooling

$$V_r = V_r(\Delta T_{kr}) \quad (7)$$

Here

$$\Delta T_{kr} = T_{er}|_{boundary} - T_r|_{boundary} \quad (8)$$

is kinetic overcooling, $T_{er}$ is the equilibrium temperature of the phase transition. Expand the velocity by kinetic overcooling into a Maclaurin series

$$V_r = V_r|_{\Delta T_{kr}=0} + \frac{\partial V_r}{\partial \Delta T_{kr}}\bigg|_{\Delta T_{kr}=0} \Delta T_{kr} = \frac{\partial V_r}{\partial \Delta T_{kr}}\bigg|_{\Delta T_{kr}=0} \Delta T_{kr} = \Lambda_r \Delta T_{kr} \quad (9)$$

Here the kinetic coefficient is introduced

$$\Lambda_r = \frac{\partial V_r}{\partial \Delta T_{kr}}\bigg|_{\Delta T_{kr}=0}$$

From (8, 9) we find

$$T_r|_{boundary} = T_{er} + \frac{V_r}{\Lambda_r}$$

Crystallization temperature is a function of the equilibrium temperature of the phase transition for the given component concentration and interface curvature [3].

$$T_r\big|_{boundary} = T_{e0r} + m_r\left(C_r\big|_{boundary} - C_{e0r}\right) + \Gamma_r \kappa_r\left(T_r\big|_{boundary}, C_r\big|_{boundary}\right) - \frac{V_r}{\Lambda_r} \quad (10)$$

Where $\kappa_r$ is the interface curvature, $\Gamma_r$ is the surface tension coefficient. At large values of the kinetic coefficient the last term becomes small and the equation takes the form

$$T_r\big|_{boundary} = T_{e0r} + m_r\left(C_r\big|_{boundary} - C_{e0r}\right) + \Gamma_r \kappa_r\left(T_r\big|_{boundary}, C_r\big|_{boundary}\right)$$

i.e. the interface temperature is equal to the temperature of phase transition. This is the limiting case considered in the works on directed crystallization. There is a formal reason to take the interface temperature as equal to the temperature of phase transition at the nonzero velocity. Here the smallest term rather than kinetic overcooling is neglected. A different situation occurs when the dependence of interface velocity on kinetic overcooling is neglected, and interface perturbation is used instead of temperature and concentration perturbation expansion of the velocity If the kinetic overcooling in (9) is assumed to be equal to zero, then $V_r = 0$. This result conceals an error which occurs when solving nonstationary problems, in particular, in analysis of the stability of the stationary regime of interface movement. Consider expansion of velocity by kinetic overcooling in the vicinity of the stationary crystallization regime for small perturbations of stationary temperature and concentration values. Then kinetic overcooling is written as a sum of perturbation and the stationary part

$$\Delta T_{kr} = \Delta T_{krS} + \Delta T_{krm}$$

Then the linear approximation of Taylor series expansion of rate in the vicinity of the stationary value of kinetic overcooling takes the form

$$V_r \approx V_r(\Delta T_{krS}) + \frac{\partial V_r}{\partial \Delta T_{kr}}\bigg|_{\Delta T_{kr}=\Delta T_{krS}} \Delta T_{krm} = V_r(\Delta T_{krS}) + \Lambda_r \Delta T_{km} = V_{Sr} + V_{mr}$$

Here the following notations are used

$$V_S = V_r(\Delta T_{krS}) \quad (11)$$

$$V_{mr} = \Lambda_r \Delta T_{km}$$

The expansion yields

$$\Delta T_{km} = \frac{V_{mr}}{\Lambda_r}$$

For the stationary mode expansion (10) can be written as

$$T_r\big|_{boundary} = T_{e0r} + m_r\left(C_r\big|_{boundary} - C_{e0r}\right) + \Gamma_r \kappa_r\left(T_r\big|_{boundary}, C_r\big|_{boundary}\right) - \frac{V_{Sr}}{\Lambda_r} - \frac{V_{mr}}{\Lambda_r}$$

Here expansion (9) is used for function (11). For sufficiently large values of the kinetic coefficient the interface temperature is equilibrium which does not imply that kinetic overcooling is zero. In any case, formal substitution of zero kinetic overcooling leads to a zero interface velocity. When solving the stationary problem, kinetic overcooling is not introduced not due to its smallness, rather due to its constant value which is unambiguously related to the stationary interface velocity. Yet, decomposition of velocity into its stationary part and small perturbation, $V_r = V_{Sr} + V_{mr}$, without considering kinetic overcooling has no physical meaning in the problem involving perturbation of stationary concentration and temperature distribution. It does not allow for the fact that the velocity is a function of kinetic overcooling which, in its turn, is a function of phase transition and interface temperatures. At zero kinetic overcooling the equation $V_r = V_{Sr} + V_{mr}$ means that the interface is shifted with respect to the liquid phase without regard for the kinetics of particle attachment to the growing solid phase surface and the effect of the thermodynamic conditions on the liquid phase in front of the interface. As such an approach is applied to problems of linear analysis of interface stability, the initial perturbation is commonly that of the interface. Consider the physical values for the corresponding problem of small perturbations [18, 19]. In the stationary mode the interface is usually assumed to be planar and perturbation is taken as a small harmonic deviation of the interface from the stationary regime plane under the assumption that concentration and temperature deviations from the stationary solution are the results of spatial interface distortion. Then small spatial interface perturbations from the stationary solution plane are connected by the relationships in which concentration and temperature perturbations turn to zero at zero amplitude of small spatial interface perturbations. Such problem setting is fairly objectionable. If no account is taken of the dependence of interface velocity on kinetic overcooling, then no equations can be obtained revealing that small concentration and temperature perturbations cause small spatial interface perturbations (which is also referred to time pulsations). The equation describing interface velocity as a function of interface kinetic overcooling connects the temperature, concentration and spatial perturbations of the interface. From this dependence it does not follow that the interface is necessarily curved at nonzero deviations of the stationary solutions of temperature and concentration. In such problem setting the cause and effect are interchanged. Indeed, it is not the spatial deviation of the interface from the stationary solution which is responsible for the deviations of the stationary solutions of temperature and concentration, rather perturbations of the temperature and concentration stationary regime can either lead to or fail to lead to spatial perturbations of the interface [10]. Spatial perturbations of the interface are determined only by rate fluctuations caused solely by kinetic overcooling. Therefore, the quantitative description of the crystallization process should involve variation of spatial interface position only as a function of interface velocity caused by variation of kinetic overcooling.

**Consideration of interface adsorption.**

The use of the equilibrium distribution coefficient requires a more precise definition of the diffusion problem since it results in the fact that phase redistribution of components is independent of interface velocity. This is inconsistent with the experimental data. Phase transition is a nonequilibrium process and the value of the equilibrium distribution coefficient is taken from the equilibrium phase diagram. The latter is calculated at equal chemical phase potentials. The values of the chemical potentials correspond to an infinite volume of each phase. It is known, however, that component adsorption occurs on the interface [20]. Interphase adsorption results from the requirement of the equality of chemical potentials on the interface. In phase transition the interphase component redistribution is affected by the adsorption layer. The behavior of the adsorption layer upon planar interface movement was considered by Hall [21, 22]. According to Hall's theory, there is relaxation time between the component concentrations in the layer and in the solid phase. Therefore, the interface velocity must be compared not only to the diffusion constant, but also the component relaxation rate in the adsorption layer. Hall introduced an effective distribution coefficient which is equal to the ratio of component concentration in the solid phase and component concentration in liquid solution. The effective distribution coefficient was calculated by the expression

$$k_{eff} = k_0 + (k_s - k_0)\exp\left(-\frac{v_{ads}}{V_r}\right) \quad (12)$$

Here $k_{eff}$ is the effective distribution coefficient, $k_0$ the equilibrium distribution coefficient, $k_{ads}$ the equilibrium adsorption distribution coefficient, $v_{ads}$ the adsorption rate constant. The formula has the following physical meaning. To maintain the composition of the adsorption layer, the rate of component atom diffusion from the melt to the crystal must be higher, the higher the growth velocity $V_r$. This may give rise to a dissolved component concentration gradient in the melt in the direction opposite to the previous one and, as a result, a component depleted zone may form instead of an accumulation region corresponding to the equilibrium phase diagram without regard for interphase adsorption. In this case the effective distribution coefficients pass from region $k_{ads} < 1$ with the equilibrium distribution coefficient $k_0$ into the region where $k_{ads} > 1$. In the high interface velocity limit the distribution coefficient will approach the adsorption distribution coefficient, $k_{ads}$. In the quasi-equilibrium boundary conditions interphase adsorption is commonly neglected. We present linear analysis of interface stability with due regard for the distribution coefficient as a function of interface velocity (12) and derive the period of morphological instability structure.

## Model.

With allowance made for distribution of ambient temperature, heat conduction equations (1) take the form

$$\frac{\partial \overline{T}'_r}{\partial \tau_r} = \chi'_r \Delta \overline{T}'_r - \phi_r \left( \overline{T}'_r - \overline{T}_{r\,ext} \right) \quad \frac{\partial \overline{T}_r}{\partial \tau_r} = \chi_r \Delta \overline{T}_r - \phi_r \left( \overline{T}_r - \overline{T}_{r\,ext} \right) \quad (13)$$

Here $\phi$ is the heat-transfer coefficient. Mass flow balance condition (5) with due account of the effective distribution coefficient and the one-dimensionality of the problem takes the form

$$D_r \frac{\partial \overline{C}_r}{\partial z}\bigg|_{liquid} = V_r \overline{C}_r (k_{eff} - 1)\bigg|_{liquid} \quad (14)$$

Introduce dimensionless parameters into equations (2-4,13,14). To this end, multiply heat conduction equations (13) by factor $\left( \alpha^2 \chi_0 T_{e0} \right)^{-1}$, where $\chi_0 = 10^{-5} \text{м}^2 \text{с}^{-1}$, $\alpha = 10^2 \text{м}^{-1}$ are auxiliary parameters, $T_{e0r}$ is the equilibrium temperature of phase transition at original liquid solution component concentration $C_0$. Multiply diffusion equations (2) by factor $\left( \alpha^2 \chi_0 C_0 \right)^{-1}$. Multiply boundary conditions (3,4), (14) by factors $T_{e0r}^{-1}$, $\left( \alpha \chi_0 T_{e0r} \right)^{-1}$ and $\left( \alpha \chi_0 C_0 \right)^{-1}$, respectively. Interface coordinate $\overline{z}_b$ in the lab coordinate system us written as

$$\overline{z}_b(\overline{y}, \tau) = \int V(\overline{y}, \overline{z}_b, \tau) d\tau = F(\overline{y}, \overline{z}_b, \tau) \quad (15)$$

Interface velocity $V(\overline{y}, \overline{z}, \tau)$ is connected with $v_n$ by the expression

$$V = \sqrt{1 + \left( \frac{\partial F}{\partial y} \right)^2} \cdot v_n$$

Now let us write the boundary-value problem in the moving coordinate system rigidly bound to the interface. Note that the introduced coordinate system is curved with respect to the lab coordinate system. It is connected to the interface, whose velocity, in the general case, is a function of the temperature and concentration of the component, rather than to the interface moving in the stationary regime, i.e. in the lab coordinate system with a constant velocity. New variables are introduced in accordance with the expressions

$$y = \bar{y}, \quad z = \bar{z} - F(\bar{y}, \bar{z}_b, \tau)$$

In equations (1) - (5) we neglect the solid phase diffusion coefficient, and in the moving coordinate system which is rigidly bound to the interface the equations have the form

$$\frac{\partial T'}{\partial \tau} = \chi' \cdot \left[ \frac{\partial^2 T'}{\partial y^2} + \left(1 + \left(\frac{\partial F}{\partial y}\right)^2\right) \frac{\partial^2 T'}{\partial z^2} - 2\frac{\partial F}{\partial y}\frac{\partial^2 T'}{\partial y \partial z} - \frac{\partial^2 F}{\partial y^2}\frac{\partial T'}{\partial z} \right] + \frac{\partial F}{\partial \tau}\frac{\partial T'}{\partial z} - \phi(T' - T_{ext}) \quad -\infty < z \leq 0 \; (16)$$

$$\frac{\partial T}{\partial \tau} = \chi \cdot \left[ \frac{\partial^2 T}{\partial y^2} + \left(1 + \left(\frac{\partial F}{\partial y}\right)^2\right) \frac{\partial^2 T}{\partial z^2} - 2\frac{\partial F}{\partial y}\frac{\partial^2 T}{\partial y \partial z} - \frac{\partial^2 F}{\partial y^2}\frac{\partial T}{\partial z} \right] + \frac{\partial F}{\partial \tau}\frac{\partial T}{\partial z} - \phi(T - T_{ext}) \quad 0 \leq z < \infty \; (17)$$

$$\frac{\partial C}{\partial \tau} = D \cdot \left[ \frac{\partial^2 C}{\partial y^2} + \left(1 + \left(\frac{\partial F}{\partial y}\right)^2\right) \frac{\partial^2 C}{\partial z^2} - 2\frac{\partial F}{\partial y}\frac{\partial^2 C}{\partial y \partial z} - \frac{\partial^2 F}{\partial y^2}\frac{\partial C}{\partial z} \right] + \frac{\partial F}{\partial \tau}\frac{\partial C}{\partial z} \quad 0 \leq z < \infty \; (18)$$

$$T'\big|_{z=0-0} = T\big|_{z=0+0} \quad (19)$$

$$\left[1 + \left(\frac{\partial F}{\partial y}\right)^2\right]\left(\chi'\frac{\partial T'}{\partial z}\bigg|_{z=0-0} - \chi\frac{\partial T}{\partial z}\bigg|_{z=0+0}\right) - \frac{\partial F}{\partial y}\left(\chi'\frac{\partial T'}{\partial y}\bigg|_{z=0-0} - \chi\frac{\partial T}{\partial y}\bigg|_{z=0+0}\right) = \varepsilon V \quad (20)$$

$$D\left(-F'_y\frac{\partial C}{\partial y} + \left(1 + (F'_y)^2\right)\frac{\partial C}{\partial z}\right)\bigg|_{z=0+0} = (k_{eff} - 1)CV\big|_{z=0+0} \quad (21)$$

These conditions should be supplemented with specified temperature in infinitely remote points

$$T'\big|_{z \to -\infty} = T_{ext}(-\infty) \quad T\big|_{z \to \infty} = T_{ext}(\infty) \quad (22)$$

The interface velocity as a function of kinetic overcooling (7) is also specified. Let the melt overcooling conditions be such that in the stationary regime the planar crystallization front moves in the lab coordinate system with constant rate $V_S$. We study the stability of the stationary crystallization mode to small temperature and concentration perturbations in the linear approximation. To obtain the linear approximation of boundary problem (16)-(22) we assume that the solutions take the form

$$T' = T'_S(z) + T'_m(z)\exp(\omega\tau + Ky) = T'_S(z) + f'_T(z, y, \tau)$$

$$T = T_S(z) + T_m(z)\exp(\omega\tau + Ky) = T_S(z) + f_T(z, y, \tau)$$

$$C = C_S(z) + C_m(z)\exp(\omega\tau + Ky) = C_S(z) + f_c(z, y, \tau)$$

$$\omega = \omega_1 + i\omega_2; \quad K = K_1 + iK_2 \quad (23)$$

$$T'_m(z) \ll T'_S(z); \quad T_m(z) \ll T_S(z); \quad C_m(z) \ll C_S(z)$$

where

$$T'_S(z), T_S(z), C_S(z)$$

are the solutions of the stationary problem. Equation (15) for the constant rate yields

$$F = V_S \tau$$

Boundary problem (16)-(22) for stationary concentration and temperature distribution takes the form

$$\chi' \cdot \frac{\partial^2 T'}{\partial z^2} + V_S \frac{\partial T'}{\partial z} - \phi(T' - T_{ext}) = 0, \quad \chi \cdot \frac{\partial^2 T}{\partial z^2} + V_S \frac{\partial T}{\partial z} - \phi(T - T_{ext}) = 0$$

$$T'_S\big|_{z=0-0} = T_S\big|_{z=0+0}, \quad \chi' \frac{\partial T'_S}{\partial z}\bigg|_{z=0-0} - \chi \frac{\partial T_S}{\partial z}\bigg|_{z=0+0} = \varepsilon V_S, \quad T'_S\big|_{z\to-\infty} = T_{ext}(-\infty) \quad T_S\big|_{z\to\infty} = T_{ext}(\infty)$$

$$D\frac{\partial^2 C}{\partial z^2} + V_S \frac{\partial C}{\partial z} = 0,$$

$$D\frac{\partial \overline{C}}{\partial z}\bigg|_{z=0+0} = (k-1)C_S V_S\big|_{z=0+0}, \quad C_S(\infty) = 1$$

Hence, the solution of the stationary problem is

$$T'_S(z) = a'\exp(g'z) + T'_{inf} - \frac{\phi(T_{inf} + x - T'_{inf})}{\frac{\chi'\phi_0^2}{(T_{inf} + x - T'_{inf})^2} + \frac{V_S\phi_0}{x} - \phi}\exp\left(\frac{\phi_0 z}{T_{inf} + x - T'_{inf}}\right)$$

$$T_S(z) = a\exp(gz) + T_{inf} - \frac{\phi x}{\frac{\chi\phi_0^2}{x^2} + \frac{V_S\phi_0}{x} - \phi}\exp\left(\frac{\phi_0 z}{x}\right)$$

Here $a$, $a'$ is the constant of integration determined from boundary conditions (24), (25), $g$, $g'$ are the characteristic numbers of homogeneous equations. In the stationary problem the interface velocity is specified which, thus, determines the interface temperature whose deviation from equilibrium temperature of phase transition is defined by the growth mechanism. Parameter *x* is found by equating the solution of the problem to this temperature. We do not present the detailed solution of the stationary problem. The numerical calculations are given in the brief communication [17]. The solution of the stationary diffusion problem takes the form

$$C(z) = 1 + \frac{1-k_{eff}}{k_{eff}} \exp(-V_S z / D)$$

This solution is different from the known one by the distribution coefficient which is here equal to the effective distribution coefficient (12). The linear approximation for small perturbations is

$$\chi' \frac{\partial^2 T'_m}{\partial z^2} + V_S \frac{\partial T'_m}{\partial z} + \left(\chi' K^2 - \omega\right) T'_m = \left[\left(\chi' K^2 - \omega\right) \frac{\partial T'_S}{\partial z} - \phi \frac{\partial T_{ext}}{\partial F}\right] \frac{V_m}{\omega} \quad -\infty < z \leq 0 \quad (26)$$

$$\chi' \frac{\partial^2 T_m}{\partial z^2} + V_S \frac{\partial T_m}{\partial z} + \left(\chi K^2 - \omega\right) T_m = \left[\left(\chi K^2 - \omega\right) \frac{\partial T_S}{\partial z} - \phi \frac{\partial T_{ext}}{\partial F}\right] \frac{V_m}{\omega} \quad 0 \leq z < \infty \quad (27)$$

$$D \frac{\partial^2 C_m}{\partial z^2} + V_S \frac{\partial C_m}{\partial z} + \left(DK^2 - \omega\right) C_m = \frac{\left(DK^2 - \omega\right)}{\omega} V_m \frac{\partial C_S}{\partial z} \quad 0 \leq z < \infty \quad (28)$$

$$\chi' \frac{\partial T'_m}{\partial z}\bigg|_{z=0-0} - \chi \frac{\partial T_m}{\partial z}\bigg|_{z=0+0} = \varepsilon V_m \quad (29)$$

$$T'_m\big|_{z=0-0} = T_m\big|_{z=0+0} \quad T'_m\big|_{z \to -\infty} = 0 \quad T_m\big|_{z \to \infty} = 0 \quad (30)$$

$$D \frac{\partial C_m}{\partial z}\bigg|_{z=0+0} = \left(k_{eff} - 1\right)\left(V_S C_m + C_S V_m\right) + V_S C_S \frac{\partial k_{eff}}{\partial V} V_m \quad C_m\big|_{z \to \infty} = 0 \quad (31)$$

$$C_m\big|_{z \to \infty} = 0 \quad (32)$$

Let interface velocity as a function of kinetic overcooling (7) be written as

$$V = V\left(\Delta T_k\left(T(0, y, \tau), C(0, y, \tau), \kappa\right)\right) \quad (33)$$

Consider linearization of kinetic equation (33). It describes the dependence of the interface velocity on the kinetics of molecule attachment to the growing surface. The form of dependence

$V(\Delta T_k)$ is set by the model of mechanism growth. In accordance with the above change of variables, kinetic overcooling (8) is given by the expression

$$\Delta T_k = 1 + m(C(0,y,\tau)-1) + \Gamma\kappa(T(0,y,\tau),C(0,y,\tau)) - T(0,y,\tau)$$

which is general for any growth model. Expansion of rate (33) into a Maclaurin series by small temperature and concentration perturbations takes the form

$$V \approx V_S + \Lambda \cdot (-T_m + mC_m + \Gamma\kappa) \quad (34)$$

For the model of normal growth [22]

$$V = h_n \Delta T_k; \quad \Lambda_n = h_n; \quad (35)$$

For the model of screw dislocation growth

$$V = h_d \Delta T_k^2 \quad \Lambda_d = 2\sqrt{h_d v_S}; \quad (36)$$

For growth involving two-dimensional nucleation

$$V = h_1 \exp\left(-\frac{h_2}{\Delta T_k}\right) \quad \Lambda_d = \frac{V_S}{h_2}\ln^2\left(\frac{V_S}{h_1}\right) \quad (37)$$

Write linear approximation (34) by small temperature and concentration perturbations as

$$V \approx V_S + V_m = V_S + \theta f_{T0} + \gamma f_{C0}; \quad (38)$$

Here $V_m$ denotes the small perturbation of the interface velocity (rate), the small perturbations of temperature and concentration take the form

$$f_{T0} = T_m(0)\exp(Ky+\omega\tau) \quad f_{C0} = C_m(0)\exp(Ky+\omega\tau)$$

Coefficients θ and γ are expressed as

$$\theta = \Lambda \frac{\omega}{\Lambda\Gamma K^2 - \omega} \quad (39)$$

$$\gamma = -m\Lambda \frac{\omega}{\Lambda\Gamma K^2 - \omega} = -m\theta \quad (40)$$

that are found using the interface curvature expansion by small perturbations of temperature and concentration. A similar calculation was made in [13]. The dispersion equation is also found as in [17]. Find the solution of problem (26) – (32), (38). On the interface the solution yields a linear system of equations with respect to coefficients $T_{m0} = T_m(0)$ и $C_{m0} = C_m(0)$

$$(S'_T - S_T)T_{m0} - \eta V_{m0} = 0 \quad (41)$$

$$[S + 2(1-k)]C_{m0} - \left[(k-1)\xi + \frac{2k_V}{k}\right]V_{m0} = 0 \quad (42)$$

(41) is the solution of the heat conduction equation, (42) is the solution of the diffusion problem. Here $S'_T$, $S_T$, $S$ are the roots of the characteristic equations for equations (26) – (28), $\eta$ and $\xi$ are dependent on the parameters of the system with $\xi \neq 0$ at k = 1. Substituting (38) in (41), (42), we obtain a dispersion equation in the form

$$\left[(S'_T - S_T) - \eta\theta\right]\left[S + (1-k)(2+\xi\gamma) + \frac{2k_V\gamma}{k}\right] + \left[(1-k)\xi - \frac{2k_V}{k}\right]\eta\gamma\theta = 0 \quad (43)$$

where $\eta$ and $\gamma$ are the parameters depending on time frequency and the wave number. The following notation is also introduced

$$k_V = \left.\frac{dk_{eff}}{dV}\right|_{V=V_S}$$

To obtain an analytical solution for the instability period, consider dispersion equation (43) at zero frequency $\omega_2 = 0$. In this case all the parameters of the dispersion equation are real numbers. We consider a stationary mode with $k_{eff} \approx 1$, but with fulfilled inequalities $k_{eff} > 1$ and $k_V \ll 1$. Let the dispersion equation be written as

$$S'_T - S_T - \left(1 + \frac{2k_V\gamma}{S + 2k_V\gamma}\right)\eta\theta = 0$$

The fraction in the brackets is expanded into series by the small parameter

$$S'_T - S_T - \left(1 + \frac{2k_V\gamma}{S}\right)\eta\theta = 0 \quad (44)$$

For further simplifications the results of numerical calculations are required. To this end, we specify the values of the segregation coefficient k = 1.03 and liquidus slope m = - 0.05 as in [9-12], where the value of k modeled the melt layering phenomenon on the interface. Assume the external field temperature gradient be $\phi_{0r} = 10^4 \text{K m}^{-1}$. As in [11], for the numerical calculation we use the screw dislocation growth model with kinetic coefficient h = 2.2·10$^{11}$, setting Γ = 10$^{-5}$. $\eta$ can be expressed as

$$\eta = \frac{2\varepsilon}{V_S}(1+\beta)$$

As shown by the numerical calculations made in [17], $\beta \ll 1$. Substitution of this condition in (44) yields

$$S'_T - S_T - \frac{2\varepsilon}{V_S}\left(1 + \frac{2k_V \gamma}{S}\right)\theta = 0$$

Characteristic numbers $S'_T$ и $S$ take the form

$$S'_T = -1 + \sqrt{1 + \frac{\chi'^2 Y}{D^2} + \frac{\chi' \delta}{D}} \quad S_T = -1 + \sqrt{1 + \frac{\chi^2 Y}{D^2} + \frac{\chi \delta}{D}} \quad S = -1 + \sqrt{1 + Y + \delta} \quad (46)$$

where

$$\delta = \frac{4D\omega_1}{V_S^2}; \quad Y = \frac{4D^2 K_2^2}{V_S^2},$$

Substitution of the numerical values leads to the relationships

$$1 \ll \frac{4\chi' \phi}{V_S^2} \ll \frac{\chi' \delta}{D} \ll \frac{\chi'^2 Y}{D^2}$$

$$1 \ll \frac{4\chi \phi}{V_S^2} \ll \frac{\chi \delta}{D} \ll \frac{\chi^2 Y}{D^2}$$

$$1 \ll \delta \ll Y$$

which bring the characteristic numbers (46) to the form

$$S'_T = -1 + \frac{\chi'}{D}\sqrt{Y} \quad S_T = -1 - \frac{\chi}{D}\sqrt{Y} \quad S = -1 - Y$$

Substituting the expressions for the characteristic numbers in dispersion equation (45)

$$(\chi' + \chi)\frac{Y}{D} - 2\left(\sqrt{Y} - 2k_V \gamma\right)\frac{\varepsilon\theta}{V_S} = 0$$

Whence we find

$$\sqrt{Y} = \frac{\varepsilon\theta D}{V_S(\chi' + \chi)}\left(1 \pm \sqrt{1 - \frac{4k_V \gamma V_S(\chi' + \chi)}{\varepsilon\theta D}}\right)$$

At $k_v = 0$ the minus before the root gives the trivial value $Y=0$. Therefore, we consider the solution with a positive root. Linearize the expression by $k_v$.

$$\sqrt{Y} - \frac{2\varepsilon\theta D}{V_S(\chi'+\chi)} + 2k_V\gamma = 0 \quad (47)$$

Expression (39) is written as

$$\theta = \Lambda\left(\frac{\Lambda\Gamma Y}{\Lambda\Gamma Y + D\delta} - 1\right) \quad (48)$$

And substitute (48) and (40) in (47). Following elementary transformations, the equation obtained is written as a fraction. Equating the numerator to zero, we arrive at an equation as related to $Y$

$$(\chi'+\chi)V_S\Lambda\Gamma\left(\sqrt{Y}\right)^3 + (\chi'+\chi)DV_S\delta\sqrt{Y} + 2(D\varepsilon + k_V mV_S(\chi'+\chi))D\Lambda\delta = 0 \quad (49)$$

Instead of $\delta$, we introduce the relationship

$$\delta = NY$$

Where $N = \frac{\delta}{Y}$. According to the numerical calculation [11], at the parameters specified

$$\delta \sim Y, \quad (50)$$

Thus, $N \sim 1$ and from (49) we find

$$\sqrt{Y} = \frac{2D\Lambda N(D\varepsilon + k_V mV_S(\chi'+\chi))}{V_S(\chi'+\chi)(\Lambda\Gamma + DN)}$$

Whence we find the period of spatial perturbations

$$\lambda = \frac{2\pi(\chi'+\chi)}{\varepsilon\Lambda + \dfrac{k_V mV_S(\chi'+\chi)}{D}}\left(1 + \frac{\Lambda\Gamma}{ND}\right) \quad (51)$$

The expression obtained is distinguished from the time expression in [13] by the dependence of time on the liquidus slope and the diffusion coefficient. In the limit $k_v = 0$ we obtain the expression of [13].

Condition (50) is essential for obtaining the desired solution. The validity of this condition follows from the numerical calculations of the solutions of the system dispersion equation [11,12]. Substituting expressions (35) - (37) in (51), for normal growth model we find

$$\lambda_n = \frac{2\pi(\chi'+\chi)}{\varepsilon h_n + \frac{k_v m V_S(\chi'+\chi)}{D}}\left(1+\frac{\Gamma h_n}{D}\right)$$

It should also be noted that in contrast to the case when interface diffusion is neglected, in normal growth the time depends on the rate of the stationary mode. For the screw dislocation growth model

$$\lambda_d = \frac{2\pi(\chi'+\chi)}{2\varepsilon\sqrt{h_d V_S} + \frac{k_v m V_S(\chi'+\chi)}{D}}\left(1+\frac{2\Gamma\sqrt{V_S h_n}}{D}\right)$$

The growth model for two-dimensional nucleation yields

$$\lambda_e = \frac{2\pi(\chi'+\chi)}{\varepsilon V_S h_2 \ln\left(\frac{V_S}{h_1}\right)^2 + \frac{k_v m V_S(\chi'+\chi)}{D}}\left(1+\frac{\Gamma V_S h_2}{D}\ln\left(\frac{V_S}{h_1}\right)^2\right)$$

The expressions obtained for the morphological instability period are distinguished from the similar dependences in [13] by the presence of the parameters of the diffusion problem. The numerical calculations [11,12] showed that the period is dependent on the liquidus slope and the diffusion coefficient, though these parameters do not enter the period expression.

**Discussion and conclusions.**

The accomplished analysis of the boundary conditions reveals the reasons for interface instability and enables their simple physical explanation. Interface instability indicates that at small concentration or temperature perturbations the interface velocity increases and (in the linear approximation) tends to infinity. Let us outline the reasons for interface instability. Kinetic overcooling is driving force of the crystallization. The interface is moveless if the kinetic overcooling is zero and the interface velocity increases monotonically with increasing kinetic overcooling. The latter changes for two reasons: on changing the equilibrium temperature of phase transition or interface temperature. One of the reasons of instability is the change of the temperature of phase transition due to the changing concentration on the interface caused by interface adsorption. According to Hall [21,22], to maintain the composition of the adsorbed layer, the component atoms must diffuse from the melt to the crystal the more quickly, than interface velocity is more. Hence, a sufficiently high velocity of interface movement may give rise to a concentration gradient of the dissolved component in the melt in the direction opposite to that corresponding to low velocity. The change of the gradient direction corresponds to the transition of the effective distribution coefficient through unity. In this case a component-depleted region, rather than an accumulation region corresponding to the equilibrium phase diagram, is formed in front of the crystal. The values of the effective distribution coefficient will pass from the region $k_{\text{eff}} < 1$ with equilibrium distribution coefficient $k_0$ into the range of values $k_{\text{eff}} > 1$. In this range the increasing stationary interface velocity leads to decreasing component concentration on the interface, increasing equilibrium temperature of phase transition and, as a result, increasing kinetic overcooling and further increase of the interface rate. This instability can be illustrated by a simple diagram. Consider the distribution of concentration at k > 1 and m < 0. The diagram of instability occurrence can be shown as

$$C(0) \uparrow \Rightarrow T_e \downarrow \Rightarrow \Delta T_k \downarrow \Rightarrow V \downarrow \Rightarrow C(0) \uparrow$$

Let the stationary phase transition regime proceed in the system. Assume that with constant $T(0)$ the concentration on the interface $C(0)$ increases by $\Delta C(0)$. Since in this case $k > 1$ and $m < 0$, the equilibrium temperature of the liquid phase transition on the interface becomes somewhat smaller along with the kinetic overcooling. This also involves a decrease in the interface velocity and, hence, an increased concentration of liquid on the interface. Therefore, the initial increase of concentration leads to its further increase. The system is unstable. In the interface velocity region, where $k_{\text{eff}} < 1$, the component concentration on the interface increases with increasing interface rate, the equilibrium temperature of phase transition decreases which results in decreasing kinetic overcooling and interface velocity. The system is stable.

     The well-known concentration instability related to the so-called concentration overcooling is caused by a simultaneous change of equilibrium temperature of phase transition and interface temperature. If the temperature gradient of the liquid solution on the interface is less than the gradient of the equilibrium temperature of phase transition, the interface stability may fail. This can be schematically described as follows. Let interface temperature $T(0)$ decrease by virtue of fluctuations. Decreasing temperature involves an increase in kinetic overcooling and, hence, an increase of the interface velocity. This manifests itself as a "ridge" occurring in the region of concentration overcooling, i.e. the equilibrium temperature of phase transition on the interface gets higher. This change involves a further increase of the kinetic overcooling and the interface velocity. Thus, the region of concentration overcooling brings about interface instability. On the other hand, at $k_{\text{eff}} < 1$, in the situation in question, the increase in the interface velocity leads to increasing component concentration on the interface and, as a consequence, to decreasing kinetic overcooling and interface velocity. These two opposite processes can be illustrated by a diagram.

$$T_0 \downarrow \Rightarrow \Delta T_k \uparrow \Rightarrow v \uparrow \quad \begin{array}{l} \Rightarrow T_e \uparrow \Rightarrow v \uparrow \\ \Rightarrow C_0 \uparrow \Rightarrow T_e \downarrow \Rightarrow v \downarrow \end{array}$$

We have obtained opposite changes of equilibrium temperature of phase transition. On the one hand, it increases due to a local movement of the interface upon temperature fluctuation, on the other hand, it decreases due to edging of the component by the interface and the change of kinetic overcooling. The two opposite processes can lead to or fail to lead to interface instability which depends on the external conditions as well as the physical parameters of the system.

In conclusion, an additional comment should be made on the setting of liquid solution crystallization problems. In [11-13] the obtained dependence of the component distribution period on the interface on the interface velocity was used to explain the dependence of the eutectic structure period on the crystallization rate. This dependence coincided with the experimental data. However, it does not seem possible to relate interface stability to liquid solution component decay in the stable mode using the mechanism of displacement of one of the components by the growing interface. Component redistribution becomes apparent when assuming that the overcooled liquid solution in front of the interface is unstable and disintegrates into two phases corresponding to the eutectic temperature [15]. In this case the interface instability determines the period of solution decay.

The assumption as to the existence of an unstable solution layer complicates significantly the setting of the problem. Hence, two situations are feasible. The case of metastable solution is considered in the directed crystallization theory and has been analyzed in the present paper. The unstable solution in the non-equilibrium layer exhibits outward diffusion and tends to decay into equilibrium zones. This creates a region where the diffusion of the components differs from their diffusion in solid and liquid quasi-equilibrium solutions. In such a problem there occurs an additional interface separating the non-equilibrium and quasi-equilibrium liquid phases. Then the component distribution is described by three rather than two diffusion equations. As a result, the equilibrium phase diagram cannot be used to determine the values of concentration in the overcooled layer, or, more precisely, the condition of equality of chemical potentials is not valid on the solution interface. Therefore, one has to apply the conditions connecting the parameters of the system on the basis of the dynamics of the physical process. All the phases differ in component concentrations and diffusion coefficients. The analysis of the problem is outlined in [24,25]. The setting and the solution are based on thermodynamics of multicomponent solutions involving a concept of osmotic pressure. The latter is caused by the difference in the mobility of the components [26] which occurs on the interfaces in question. To obtain the boundary conditions for interfaces with non-equilibrium solutions, it is sufficient to abandon the mobility equality condition, i.e. the component diffusion coefficients. The idea is not new and has been used to explain the Kirkendall effect in the known experiments on atomic plane displacement in different contacting solid materiales [26]. In the one-dimensional case the stationary diffusion problem consists of three equations whose solutions are determined by six integrations constants. Four of them are defined by the following boundary conditions.

1. Specified concentration in infinitely remote point of quasi-equilibrium solution.
2. Zero diffusion coefficient in solid phase.
3. Equality of constant concentration plane velocity on both sides of solid phase-non-equilibrium liquid interface.
4. Equality of constant concentration plane velocity on both sides of non-equilibrium liquid-quasi-equilibrium liquid interface.

The other two constants are found from the condition of mass component flow conservation. Besides these conditions, the concentration in the solid phase and the quasi-equilibrium phase concentration on the non-equilibrium solution interface are bound by an equilibrium phase diagram which determines the coordinate of the interface between the non-equilibrium and quasi-equilibrium phases. The problem has an analytical solution at a small difference between the component diffusion coefficients [24,25]. The analysis of the problem is beyond the scope of this paper.

## Conclusions

1. The free interface problem for solution phase transitions has been considered for the case of directed crystallization of two-component solution. Analysis has been made of the edge conditions that have a qualitative effect on the solution of the problem.
2. An internal source field model which simulates external heat exchange forming the temperature field of crystallized solution has been suggested.
3. An analytical expression has been obtained for the dependence of the period of morphological interface instability. The expression takes into account the effect of interface adsorption on the coefficient of component distribution between liquid and solid solutions.
4. Account has been taken of the parameters of the diffusion problem, in contrast to the results obtained in [13], where the period dependences were obtained without regard for the influence of the adsorption effect.

## Reference.